\documentclass[aps, prl, reprint, superscriptaddress, showpacs, showkeys, twocolumn, bibnotes, nofootinbib]{revtex4}
\usepackage[latin10]{inputenc}
\usepackage[USenglish]{babel}
\usepackage{color}
\usepackage{amstext}
\usepackage{amsmath}
\usepackage{amssymb}
\usepackage{graphicx}
\usepackage{textcase}
\usepackage{csquotes}
\usepackage{verbatim}
\usepackage{epstopdf}
\usepackage{url}
\usepackage{hyperref}
\usepackage{natbib}
\hypersetup{colorlinks=true,linkcolor=blue,citecolor=blue,urlcolor=blue}

\begin{document}

\title{Thermally-induced crossover from 2D to 1D behavior in an array of atomic wires: silicon dangling-bond solitons in Si(553)-Au}

\author{B. Hafke}
\email[]{bernd.hafke@uni-due.de}
\thanks{Corresponding author}
\author{C. Brand}
\author{T. Witte}
\author{B. Sothmann}
	\affiliation{Faculty of Physics and Center for Nanointegration (CENIDE), University of Duisburg-Essen, 47057 Duisburg, Germany}
\author{M. \surname{Horn-von Hoegen}}
	\affiliation{Faculty of Physics and Center for Nanointegration (CENIDE), University of Duisburg-Essen, 47057 Duisburg, Germany}
 \author{S.\,C. Erwin}
	\affiliation{Center for Computational Materials Science, Naval Research Laboratory, Washington, DC 20375, United States}

\date{\today}

\begin{abstract}
The self-assembly of submonolayer amounts of Au on the densely stepped Si(553) surface creates an array of closely spaced ``atomic wires'' separated by 1.5 nm. At low temperature, charge transfer between the terraces and the row of silicon dangling bonds at the step edges leads to a charge-ordered state within the row of dangling bonds with $\times 3$ periodicity.  Interactions between the dangling bonds lead to their ordering into a fully two-dimensional (2D) array with centered registry between adjacent steps. We show that as the temperature is raised, soliton defects are created within each step edge. The concentration of solitons rises with increasing temperature and eventually destroys the 2D order by decoupling the step edges,  reducing the effective dimensionality of the system to 1D. This crossover from higher to lower dimensionality is unexpected and, indeed, opposite to the behavior in other systems.
\end{abstract}


\maketitle


Physical phenomena associated with low dimensionality are 
suppressed when the temperature is raised.
For example, the 2D fractional quantum
Hall effect \cite{Tsui:PRL48,Laughlin:PRL50} and the 1D
Tomonaga-Luttinger liquid
\cite{Tomonaga:PoTP5,Luttinger:JoMP4,Bockrath:Nature397} are only
observed at low temperature. In 1D atomic wire systems at low
temperatures, Peierls distortions or more general symmetry breakings
can open a gap at the Fermi level and lower the total energy by
forming a charge density wave (CDW)
\cite{AhnYeom:PRL93,Frigge:Nature544,SnijdersWeitering:RevModPhys82}
or spin-density wave (SDW)
\cite{Gruener:RevModPhys66,Andrieux:JPL42,Mortensen:SSC44,Sassa:JoESaRP224}. 

Excitations
generally wash out the effects of this anisotropy and
hence suppress low-dimensional behavior.
The resulting crossover to higher dimensionality at increased
temperatures is exhibited by many systems.  Recent examples include
the atomic wire systems Pt(110)-Br and Si(557)-Pb. In these systems,
structural changes are accompanied by a delicate interplay between CDW
correlations and short-range interactions of the adsorbate atoms
\cite{Duerrbeck:PRB98} and by correlated spin-orbit order that
triggers a metal-to-insulator transition, respectively \cite{Brand:NatComm6,Das:JPCM28,Tegenkamp:PRL95,Block:PRB84}. The
resulting dimensional crossover from 1D to 2D is typical for atomic
wire systems.

In this Letter we demonstrate the opposite case: a system of coupled
atomic wires exhibiting 2D order at low temperatures in which
thermal excitations at higher temperatures induce a dimensional
crossover to 1D behavior. We identify the mechanism driving this
crossover to be the creation of phase solitons and
antisolitons \cite{RiceMele:PRL49,SuSchrieffer:PRL46}, which
leads to an reversible order-disorder transition at higher temperatures
\cite{AhnYeom:PRL95}. We track the crossover across its characteristic
temperature (approximately 100\,K) using a combination of a
quantitative high resolution spot profile analyzing-low energy
electron diffraction (SPA-LEED) study, density-functional theory (DFT)
calculations,  Monte Carlo
statistical simulations, and
an exactly solvable analytical model.

\begin{figure}[b]
	\includegraphics[width=0.96\columnwidth]{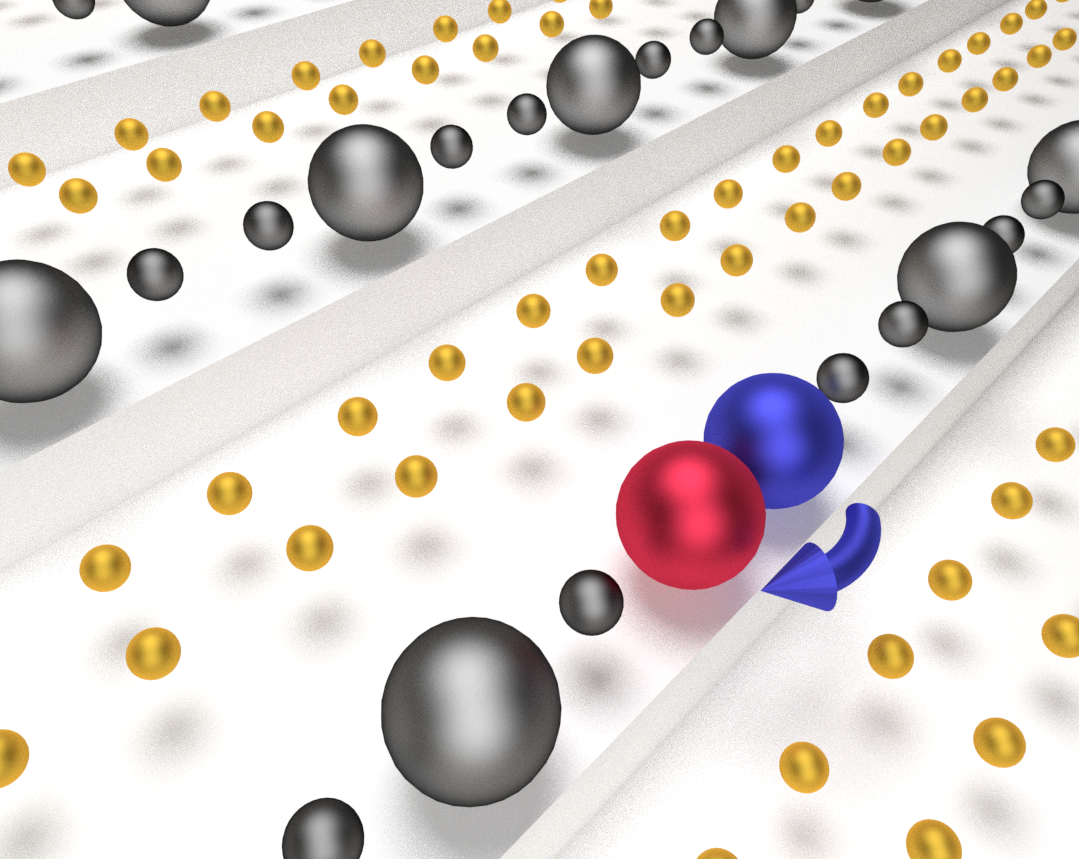}
	\caption{Ground state structure and low-energy excitation of the Si(553)-Au atomic wire system. The underlying substrate consists of Si(111) terraces separated by steps. Each terrace contains a dimerized double row of Au atoms (gold) and a row of Si dangling bonds (gray spheres) at the step edge. The electron occupancy of these dangling bonds creates a ground state with tripled periodicity: for every two saturated dangling bonds (SDBs, small spheres) there is a third, unsaturated dangling bond (UDB, large spheres). At finite temperatures, some of these UDBs (blue) become mobile and hop to adjacent sites (red). This  excitation creates a soliton-antisoliton pair that can subsequently dissociate. 
	\label{fig:Introductory}}
\end{figure}

We studied the self-organized Si(553)-Au atomic wire surface
consisting of Au double-atom rows on (111)-oriented Si terraces
separated by bilayer steps [Fig.~\ref{fig:Introductory}]. Charge
transfer from the terraces leads to incomplete filling of the dangling
$sp^3$ orbitals at the Si step edge
\cite{Crain:PRL90,SnijdersWeitering:PRL96,AhnYeom:PRL95}. The
low-temperature ground state consists of a charge-ordered state with
$\times 3$ periodicity along the step edges, which is observed in
scanning tunneling microscopy (STM)
\cite{AhnYeom:PRL95,SnijdersWeitering:PRL96,ShinYeom:PRB85,Song:ACSNano9,Aulbach:PRL111,Polei:PRL111,Aulbach:PRB96,Hafke:PRB94,Dudy:JPCM29}
and LEED experiments
\cite{Hafke:PRB94,Yeom:NJoP16,AhnYeom:PRL95,Dudy:JPCM29}. The $\times
3$ periodicity along the wires represents the simplest way to
distribute the available electrons among the row of dangling bonds
while maximizing the number of fully saturated dangling bonds
(electron lone pairs) \cite{Aulbach:NanoLett16}. Angle-resolved
photoemission spectroscopy measurements
\cite{Crain:PRL90,Song:ACSNano9,Yeom:NJoP16,Krawiec:ASS373} and DFT
calculations
\cite{Erwin:NatComm1,Song:ACSNano9,Krawiec:PRB81,Krawiec:ASS373}
reveal that the dangling-bond states do not cross the Fermi level.
Hence, all the dangling-bond orbitals have integer electron
occupancies of 0, 1, or 2.  We will refer to orbitals with
occupancy 2 as saturated dangling bonds (SDBs) and to those less than
2 as unsaturated dangling bonds (UDBs). Figure~\ref{fig:Introductory}
depicts the arrangement of UDBs and SDBs schematically.
The ordering of the Si dangling bond structure is mediated by Coulomb interaction of the UDBs (large spheres) with approximately equal spacing within and across the rows. 
	The SDBs merely provide a compensating background charge to balance the reduced electron occupancy of the UDBs.

The actual number of electrons in the UDBs has been
previously investigated using DFT. The result is sensitive to the
choice of the exchange-correlation functional. The original prediction
\cite{Erwin:NatComm1}, which used the functional of Perdew, Burke, and
Ernzerhof (PBE) \cite{Perdew:PRL77}, was that five electrons are
shared among three dangling bonds with an electron configuration
(2,2,1) having $\times 3$ periodicity and one unpaired spin
\cite{Hafke:PRB94}. More recent work \cite{Braun:PRB98}, based
on the revised PBEsol functional \cite{Perdew:PRL100}, predicted that only
four electrons are shared among three dangling bonds, implying the configuration
(2,2,0) with no unpaired spins. At present it is not possible to
distinguish between these scenarios on the basis of experimental
data. Here we consider both possibilities and show that they lead to very different estimates of the order-disorder transition temperature.

The experiment was performed under ultra-high vacuum (UHV) conditions
at a base pressure lower than $1 \times 10^{-10}$\,mbar. The Si
substrate was cut from an \textit{n}-type Si(553) wafer (phosphorus
doped, $0.01\,\Omega$\,cm). Prior to Au deposition, the sample was
cleaned in several short flash-anneal cycles by heating via direct
current to $1250\,^\circ$C. Next, 0.48\,ML (monolayer, referred
to the atomic density of a Si(111) surface, i.\,e.\ $1\,\text{ML} =
7.83 \times 10^{14}$\,atoms per cm$^2$) Au was deposited from an
electron-beam-heated graphite crucible \cite{Kury:RSI76} at a
substrate temperature of $650\,^\circ$C, followed by a post-annealing
step at $850\,^\circ$C for several seconds \cite{Aulbach:PRL111} and
subsequent cooling to 60 K on a liquid helium
cryostat. The temperature was measured by an ohmic sensor (Pt100)
directly clamped to the back of the sample.

\begin{figure}[b]
	\includegraphics[width=0.96\columnwidth]{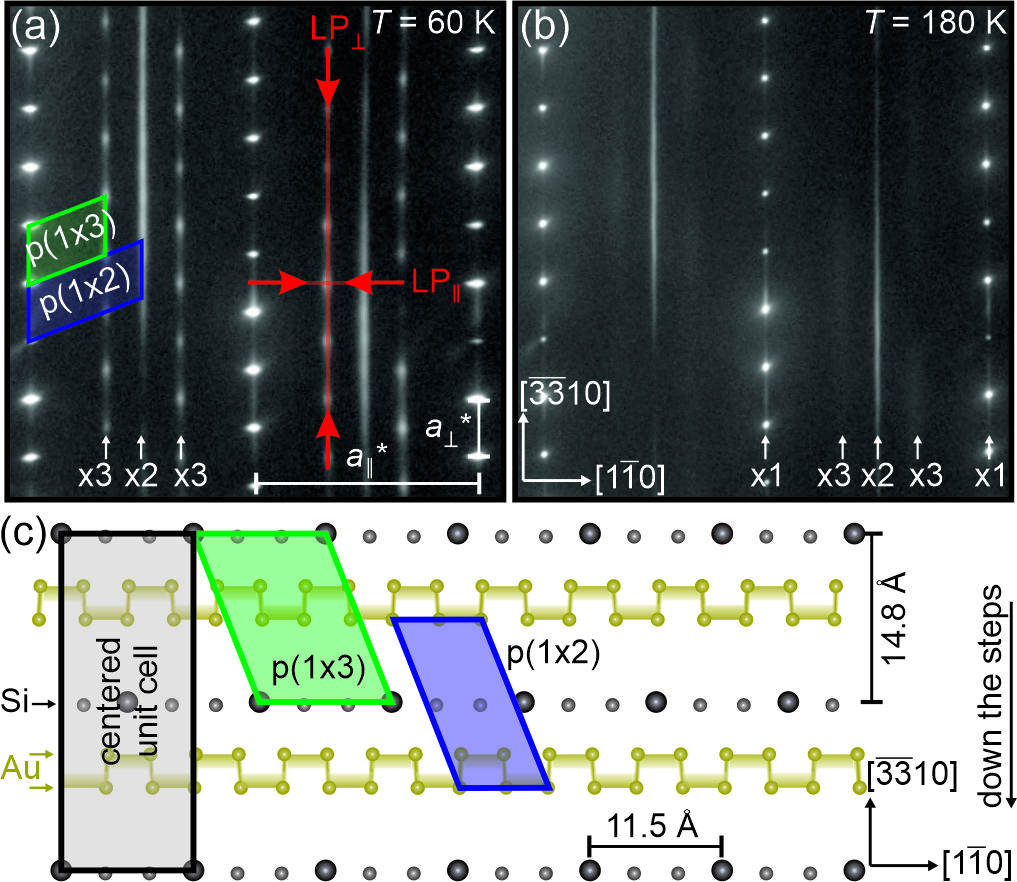}
	\caption{SPA-LEED pattern of Si(553)-Au at an electron energy
	of 150 eV and temperatures (a) 60 K and (b) 180 K. The $\times
	2$ streaks between the rows of sharp integer-order spots arise
	from dimerized Au double rows on the (111)-oriented
	terraces. The rows of elongated spots at $\times
	3$ positions indicate the tripled periodicity and long-range
	order of the UDBs at the Si step edge. The intensity of the
	$\times 2$ streak is nearly unaffected at higher temperature,
	while the $\times 3$ features fade away. In (a) the unit cells
	of the Au (blue) and Si (green) sublattices as well as the
	directions of the line profiles LP$_\parallel$ and LP$_\perp$
	(Fig.~\ref{fig:SPALEED_curves}) are indicated (for more
	details see \cite{Hafke:PRB94}). (c) Surface structural model
	showing Si step-edge atoms (gray) and Au atoms (gold). The
	unit cells are depicted with the same color coding as in
	(a).\label{fig:Diff.Pattern}} \end{figure}

\begin{figure}[t]
	\includegraphics[width=0.96\columnwidth]{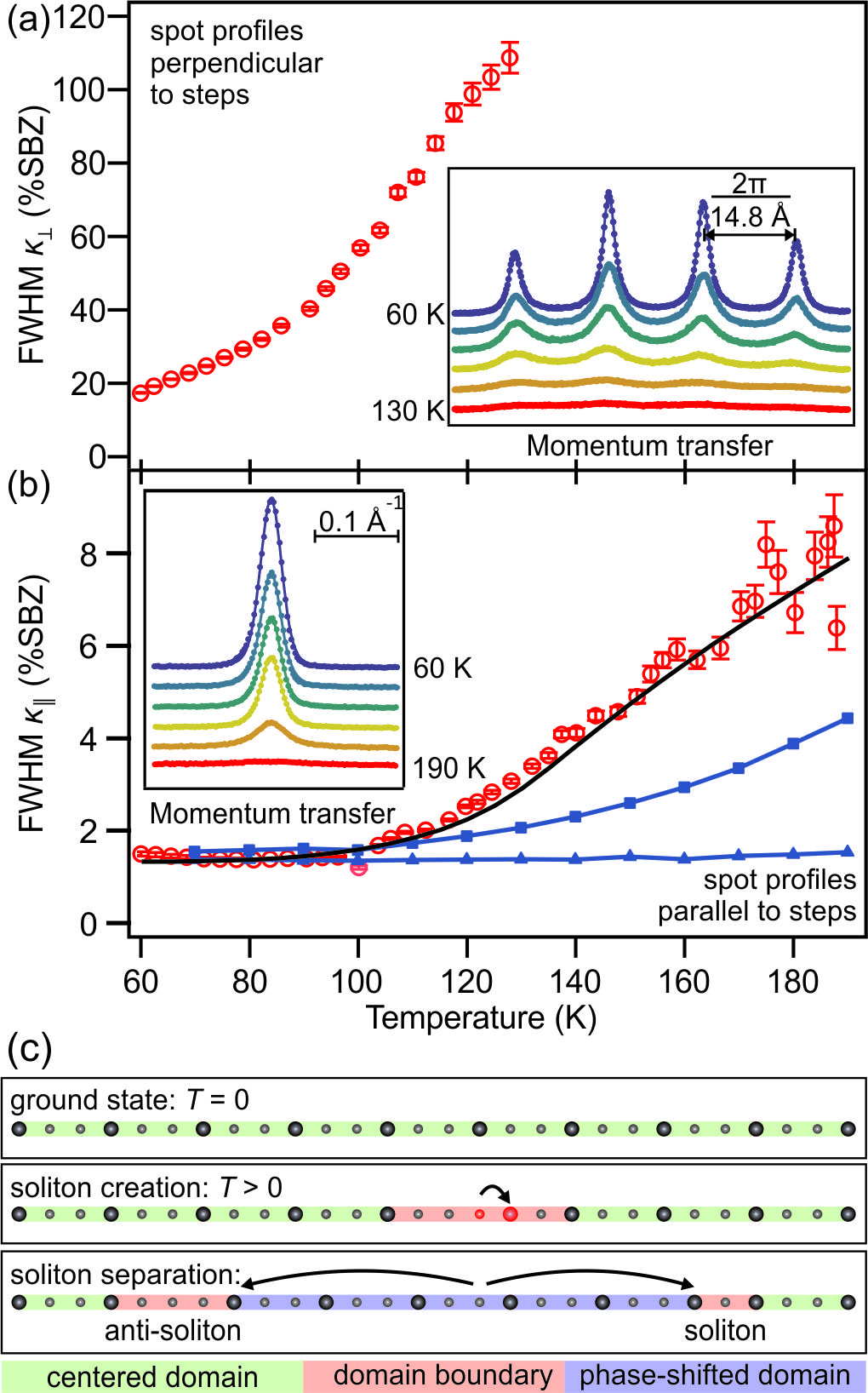}
	\caption{FWHM of the $\times 3$ diffraction spots (red data
	points) as function of temperature in (a)
	$\left[\overline{3}\,\overline{3}\,10\right]$ direction and
	(b) $\left[1\,\overline{1}\,0\right]$ direction, as indicated
	in Fig.~\ref{fig:Diff.Pattern}(a). Results from Monte Carlo
	simulations based on DFT interactions are shown in (b) by blue
	squares for the (2,2,1) configuration and triangles for the
	(2,2,0) configuration. The result from the analytical model in
	Eq.~\eqref{eq:Sothmann-Hamiltonian} is shown in black. The
	increase of the FWHM in (a) indicates loss of inter-wire
	correlation, while in (b) it indicates a decreasing
	correlation length along the steps. At low temperatures the
	FWHM is constant in
	$\left[\overline{3}\,\overline{3}\,10\right]$ direction up to
	$\sim$90 K. Insets in (a,b): Line profiles for both directions
	at various temperatures (shifted for better visibility). (c)
	Structural model of creation and separation of a
	soliton-antisoliton pair. Charge is transferred from an
	SDB to a UDB, generating a hop of the UDB to a neighboring
	site and creating a soliton-antisoliton pair. If this pair
	separates then a phase-shifted domain with $\times 3$
	periodicity is formed.\label{fig:SPALEED_curves}} \end{figure}

At 60 K, the SPA-LEED pattern [Fig.~\ref{fig:Diff.Pattern}(a)] reveals
spots at $\times 3$ positions and streaks at $\times 2$ positions in
the $\left[1\,\overline{1}\,0\right]$ direction. The latter indicates the
formation of Au atomic wires. The spacing of the $\times 1$ spots
corresponds to the reciprocal lattice constant $a_\parallel^* =
2\pi/(3.84\,\text{\AA})$ of the Si substrate along the steps.  The
$\times 3$ spots arise from ordering of the UDBs within the rows;
hence, the UDBs have an intra-row separation of $3\times a_\parallel =
11.5\,\text{\AA}$. The UDBs in different rows are in registry: recent
investigations by SPA-LEED and STM reveal a centered $p(1 \times 3)$ 
arrangement [Figs.~\ref{fig:Diff.Pattern}(a,c)]
\cite{Hafke:PRB94}. In the
$\left[\overline{3}\,\overline{3}\,10\right]$ direction the reciprocal
step-to-step distance is $a_\perp^* = 2\pi/(14.8\,\text{\AA})$
\cite{Crain:PRL90} and thus the separation between the UDB rows is 14.8
\AA. Hence, at low temperatures the UDBs in Si(553)-Au are arranged in
rows in a fully ordered 2D array with approximately equal 
spacing within (11.5~\AA) and across (14.8~\AA) the rows.
The $\times 2$ streaks are attributed to the
dimerized double row of Au atoms on the (111) terrace of the surface
[gold spheres in Fig.~\ref{fig:Diff.Pattern}(c); the unit cell is
shown by the blue-shaded areas in Figs.~\ref{fig:Diff.Pattern}(a,c)]
\cite{Krawiec:PRB81,KrawiecJalochowski:PRB87}. We did not detect any
$\times 6$ periodicity in the $\left[1\,\overline{1}\,0\right]$
direction, indicating that the Au atoms and Si step-edge atoms are
structurally decoupled \cite{Hafke:PRB94,Aulbach:PRB96,Braun:PRB98}.

At 180 K, the intensity of the $\times 3$ spots fades markedly
[Fig.~\ref{fig:Diff.Pattern}(b)], in agreement with an earlier study
\cite{AhnYeom:PRL95}, while the intensity of the $\times 2$ streaks is
nearly unaffected.  To analyze the evolution of the diffraction
pattern between 60 K to 190 K (heating rate 0.13\,K/s), we recorded
a series of line profiles [Fig.~\ref{fig:SPALEED_curves}(a,b)] through
the $\times 3$ spots, in two orthogonal directions: 
$\left[\overline{3}\,\overline{3}\,10\right]$ (LP$_\perp$,
across the steps) and
$\left[1\,\overline{1}\,0\right]$ (LP$_\parallel$,
along the steps). The $\times 3$ diffraction spots [insets of Fig.~\ref{fig:SPALEED_curves}(a,b)] of each of the line
profiles were best fitted by a series of equidistant Lorentzian
functions. No Gaussian-like central spike is found and the positions
of the $\times 3$ diffraction spots do not shift with
temperature. Across the steps, the full width at half maximum (FWHM)
$\kappa_\perp$ steadily increases as the temperature is
raised from 60\,K to 130\,K. Eventually, the spots merge into streaks,
consistent with the vanishing of the $\times 3$ periodicity reported
earlier \cite{AhnYeom:PRL95}. 

Along the steps, the FWHM $\kappa_\parallel$ is relatively constant up
to about 100 K and then steadily increases as the temperature is
raised further.  This broadening of the $\times 3$ diffraction spots
is due to increasing disorder in the arrangement of UDBs.  This type
of disorder originates from a simple microscopic process in which 
an electron (or two electrons, for the (2,2,0) configuration) hops
from an SDB onto a neighboring UDB [middle panel of
Fig.~\ref{fig:SPALEED_curves}(c)]. As long as these electron hops do
not bring neighboring UDBs closer than $2a_\parallel$, the
configuration is metastable.

We used DFT to determine the formation energy $E_0$ of this elementary
excitation, which can be viewed as a soliton-antisoliton bound
pair. The calculations were performed in a 1$\times$6 cell of
Si(553)Au with four silicon double layers plus the reconstructed
surface layer and a vacuum region of 10\,\AA. All atomic positions
were relaxed except the hydrogen-passivated bottom double layer.
Total energies and forces were calculated using the
generalized-gradient approximation of Perdew, Burke, and Ernzerhof
(PBE) for the (2,2,1) configuration and the PBEsol revision of that
functional for the (2,2,0) configuration, with projector-augmented
wave potentials as implemented in VASP
\cite{Kresse:PRB47,Bloechl:PRB50,Perdew:PRL100}. The plane-wave cutoff
was 250 eV and the sampling of the surface Brillouin zone was $6
\times 6$.

For the (2,2,1) ground-state configuration we find $E_0=30$ meV, suggesting these defects will be numerous at temperatures above $\sim$300 K, which is consistent with our experimental data. For the (2,2,0) configuration we find $E_0=85$ meV, implying a much higher temperature scale of $\sim$1000 K. 

To investigate the concentration and distribution of defects as a
function of temperature, we used the Metropolis Monte Carlo method to
sample the steady-state arrangement of UDBs in an infinite array of dangling-bond
wires with the Si(553)-Au geometry. We performed 10$^7$ trial hops at
each temperature and computed the diffraction intensity from the
positions of the UDBs. The spectra were convolved with a Gaussian to
account for instrumental broadening in the experimental data. For the
(2,2,1) configuration, the resulting FWHM of the $\times$3 peaks is
constant up to $\sim$100 K and then increases gradually with
temperature, in agreement with experiment but with smaller values
[blue squares in Fig.~\ref{fig:SPALEED_curves}(b)]. For the (2,2,0) configuration the FWHM
is flat up to temperatures about three times higher (blue triangles),
as expected from the $q^2$ scaling of the Coulomb energy. See
Supplementary Material for additional details.

Even though the geometry of our model is 2D, the energy scale of
Coulomb interactions across the wires is only 0.1 meV, three orders of
magnitude smaller than $E_0$. Hence, the interactions in the Monte
Carlo simulations are essentially 1D. Our DFT calculations, however,
reveal a much stronger interaction across the wires of order 1
meV. These may arise from the interaction of strain fields from the
UDBs but other sources may contribute as well. Regardless of their
origin, we turn now to investigating their role in the order-disorder
transition. We show next that by including these 2D interactions, the
FWHM at all temperatures is brought into quantitatively excellent
agreement with experiment.

We constructed an exactly solvable Potts model Hamiltonian describing
the dynamics of coupled wires and the resulting steady-state FWHM of
the $\times$3 peaks as a function of temperature: \begin{equation}
\label{eq:Sothmann-Hamiltonian} \mathcal{H} = \sum_i
\left[-b\delta_{u_i,u_{i+1}}-a\delta_{u_i,\text{c}}\right],
\end{equation} where $\delta_{i,j}$ denotes the Kronecker delta. A
single UDB can take three positions within each unit cell $i$: left,
center and right, $u_i=\{\text{l},\text{c},\text{r}\}$. The first
term, with parameter $b$, describes the energy needed to displace
neighbouring UDBs relative to each other: specifically, the energy
needed to create a soliton-antisoliton pair within one wire is
$2b$. The second term, with parameter $a$, favors the occupation of
the central position and arises from the coupling of the wire to
neighboring wires. See Supplementary Material for additional details.

The model fits best to our experimental data for $a = 2.1$ meV and $b
= 21$ meV. These fitted values are also consistent with our DFT
results: $a$ should be equal to the calculated energy difference per
UDB, $2.1$ meV, between (2,2,1) configurations in staggered and
centered alignments, and $b$ corresponds to $E_0/2 = 15$ meV.  In the
Supplementary Material, we derive analytical expressions for the
profiles and FWHM of the $\times 3$ peaks as a function of
temperature. The resulting FWHM, convolved as above with a Gaussian,
is now in excellent agreement with our experimental results [black
curve in Fig.~2(b)]. This improved agreement demonstrates that 2D
coupling between neighboring wires indeed plays an important, central
role in the order-disorder transition.

We turn now to the crossover from 2D to 1D behavior. At temperatures above $\sim$120 K, the $\times3$ diffraction spots are well described by a standard Lorentzian. At temperatures below $\sim$90 K, the 2D character of the diffraction is more pronounced and hence the spot profiles are described by a Lorentzian to the power 3/2 \cite{Wollschlaeger:PRB44}. To characterize the transition between these two limits, we fit the spot profiles to a generalized Lorentzian,
\begin{equation}
\label{eq:Lorentzian}
\mathcal{L}(k_\parallel) = \frac{\Gamma(\nu)}{\sqrt{\pi} \Gamma(\nu - 1/2)} \cdot \frac{\kappa_\parallel^{2\nu - 1}}{\left[ (k_\parallel - k_0)^2 + \kappa_\parallel^2 \right]^\nu},
\end{equation}
where $k_\parallel$ is the reciprocal space coordinate in the $\left[1\,\overline{1}\,0\right]$ direction, $k_0$ is the position of the $\times 3$ diffraction spot, and $\Gamma(x)$ is the Gamma function. The parameter $\nu = (d + 1)/2$ characterizes the dimensionality $d$ of the system: $\nu = 3/2$ describes 2D systems while $\nu = 1$ describes 1D systems  \cite{Wollschlaeger:PRB44,Lent:SurfSci139,Pukite:SurfSci161};  we constrained $\nu$ to lie in this range.

\begin{figure}[t]
	\includegraphics[width=0.96\columnwidth]{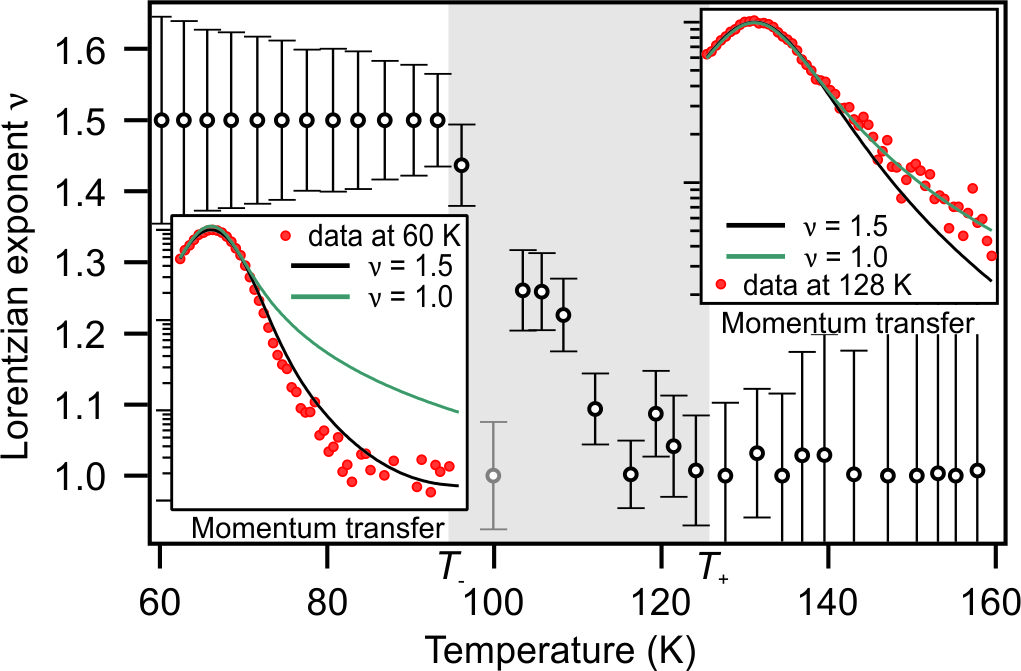}
	\caption{Temperature dependence of the exponent $\nu$ of the generalized Lorentzian of Eq.~\eqref{eq:Lorentzian} in the $\left[1\,\overline{1}\,0\right]$ direction. The exponent drops from $1.5$ at $T_{-}=93$\,K to 1.0 at $T_{+}=128$\,K, indicating a crossover from 2D to 1D behavior. Insets: Experimental and fitted spot profiles at 60 K and at 128 K. At 60 K the profile is best fit by $\nu = 3/2$ (2D behavior), while at $T_{+}$ the best fit is $\nu = 1$ (1D behavior).\label{fig:2d-1d-transition}}
\end{figure}

We find that $\nu$ exhibits a well-defined transition from 1.5 to 1.0
between $T_{-}=93$\,K and $T_{+}=128$\,K
(Fig.~\ref{fig:2d-1d-transition}). The transition begins at about the
temperature for which the FWHM $\kappa_{\parallel}$ along the steps
begins to increase [Fig.~2(b)].  Fitting the spot profiles without
allowing $\nu$ to vary leads to significantly worse fits (insets to
Fig.~\ref{fig:2d-1d-transition}). The transition is completed at
$T_{+}$, where the FWHM $\kappa_\perp$ across the wires exceeds the
size of the surface Brillouin zone [Fig.~2(a)], reflecting the
complete loss of long-range order across the wires. The
underlying origin of this dimensional crossover is subtle but simple:
the approximate geometrical isotropy of the 2D array of UDBs is broken
by the strong anisotropy of the energy scales for creating disorder
across and within the UDB wires. At temperatures above $T_{-}$ soliton
defects are still rare, but a rapidly growing fraction of the wire
rows undergoes registry shifts with respect to each other and hence the
2D low-temperature state begins to behave like a collection of
uncoupled 1D wires. As the temperature approaches $T_{+}$ this
crossover becomes nearly complete. See the Supplementary Material for
additional discussion, modeling, and analysis.

To summarize, we have shown that silicon dangling-bond solitons in
Si(553)-Au are created by thermal excitation. These defects interact
via Coulomb forces within each step-edge atomic wire and via another
mechanism, probably strain, across the wires. As the temperature is
raised, the resulting disorder destroys the $\times$3 positional
long-range order of the UDBs within each wire and their registry
across the wires. The nature of the interactions and their respective
energy scales create a dimensional crossover of the order-disorder
transition from 2D at low temperature to 1D at high temperature. The
generality of this crossover can readily be investigated---both
experimentally and using our theoretical methods---in other atomic
wire systems in the Ge/Si($hhk$)-Au family, where differences in the
surface morphology and ground-state electron configuration may lead to
further expanding our understanding of low-dimensional systems.

\begin{acknowledgments}
We acknowledge fruitful discussions with J.\ Aulbach, F.\ Hucht, J.\ K\"{o}nig and J.\ Sch\"{a}fer. This work was funded by the Deutsche Forschungsgemeinschaft (DFG, German Research Foundation) -- Projektnummer 278162697 -- SFB 1242 and through Projektnummer 194370842 -- FOR1700. B.\,S.\ acknowledges financial support from the Ministry of Innovation NRW via the ``Programm zur F\"{o}derung der R\"{u}ckkehr des hochqualifizierten Forschungsnachwuchses aus dem Ausland''. This work was partly supported by the Office of Naval Research through the Naval Research Laboratory's Basic Research Program (SCE). Computations were performed at the DoD Major Shared Resource Center at AFRL.
\end{acknowledgments}

\end{document}